\documentclass[preprint2,times]{aastex6}

\usepackage{natbib}
\usepackage{amsmath}
\usepackage{amssymb}
\usepackage{subfigure}
\usepackage{url}
\usepackage{CJK}

\renewcommand{\vec}[1]{ {\mathbf #1} }

\newcommand{\Fig}{{Figure}}

\newcommand{\SDO}{{\it SDO}}
\newcommand{\Twm}{{$|T_w|_{\rm max}$~}}
\providecommand{\dodoi}[1]{doi:~\href{http://doi.org/#1}{\nolinkurl{#1}}}
\providecommand{\url}[1]{\href{#1}{#1}}
\providecommand{\doeprint}[1]{\href{http://ascl.net/#1}{\nolinkurl{http://ascl.net/#1}}}
\providecommand{\doarXiv}[1]{\href{https://arxiv.org/abs/#1}{\nolinkurl{https://arxiv.org/abs/#1}}}

\shorttitle{Pre- to Post-flare Magnetic Flux Ropes}
\shortauthors{Duan et al.}

\begin{document}
\begin{CJK*}{UTF8}{gbsn}

\title{Variation of Magnetic Flux Ropes Through Major Solar Flares}

\author{Aiying Duan \altaffilmark{1,2}$^*$,
  Chaowei Jiang \altaffilmark{3}, 
  Zhenjun Zhou \altaffilmark{1, 4},
  Xueshang Feng \altaffilmark{3},
  Jun Cui \altaffilmark{1,4}}

\altaffiltext{1}{School of Atmospheric Sciences, Sun Yat-sen University, Zhuhai 519000, China}

\altaffiltext{2}{ School of Space and Environment, Beihang University, Beijing, 100191, China}

\altaffiltext{3}{Institute of Space Science and Applied Technology, Harbin Institute of Technology, Shenzhen 518055, China}

\altaffiltext{4}{CAS Center for Excellence in Comparative Planetology, China}

\email{*Corresponding authors:}
\email{duanaiy@mail.sysu.edu.cn}

\begin{abstract}
  It remains unclear how solar flares are triggered and in what
  conditions they can be eruptive with coronal mass
  ejections. Magnetic flux ropes (MFRs) has been suggested as the
  central magnetic structure of solar eruptions, and their ideal
  instabilities including mainly the kink instability (KI) and torus
  instability (TI) provide important candidates for triggering
  mechanisms. Here using magnetic field extrapolations from observed
  photospheric magnetograms, we systematically studied the variation
  of coronal magnetic fields, focusing on MFRs, through major flares
  including 29 eruptive and 16 confined events. We found that nearly
  90\% events possess MFR before flare and 70\% have MFR even after
  flare. We calculated the controlling parameters of KI and TI,
  including the MFR's maximum twist number and the decay index of its
  strapping field. Using the KI and TI thresholds empirically derived
  from solely the pre-flare MFRs, two distinct different regimes are
  shown in the variation of the MFR controlling parameters through
  flares. For the events with both parameters below their thresholds
  before flare, we found no systematic change of the parameters after
  the flares, in either the eruptive or confined events. In contrast,
  for the events with any of the two parameters exceeding their
  threshold before flare (most of them are eruptive), there is
  systematic decrease in the parameters to below their thresholds
  after flares. These results provide a strong constraint for the
  values of the instability thresholds and also stress the necessity
  of exploring other eruption mechanisms in addition to the ideal
  instabilities.
\end{abstract}


\section{Introduction}
\label{sec:intro}
Solar eruptions, often referred to explosive phenomena in the solar
atmosphere, are leading driver of disastrous space weather. The two
commonest forms of solar eruptions are solar flares and coronal mass
ejections (CMEs), which are often closely related with each other (but
not always). Both of them are believed to be caused by a drastic
release of free energy stored in the complex, unstable magnetic field
in the solar corona. Prediction of such eruptions is difficult since
there are many important issues of the underlying physics remaining
unresolved. For instance, the structure of the coronal magnetic field
prior to eruption and the triggering mechanism of eruption are still
undetermined; the key factor that causes the difference between
eruptive flares (flares accompanied with CMEs) and confined flares
(flares without CMEs) is still unclear.

Competing models of solar eruptions have been proposed based on either
resistive magnetohydrodynamics (MHD), i.e., magnetic reconnection, or
ideal MHD instabilities ~\citep[e.g., see reviews of][]{Forbes2006,
  Shibata2011, Aulanier2014, Schmieder2012, Schmieder2013,
  Janvier2015}. For example, in the reconnection-based models, such as
the runaway tether-cutting reconnection~\citep{Moore2001} and the
breakout reconnection~\citep{Antiochos1999}, it is assumed that before
the flare, the magnetic field has a strongly sheared configuration,
and eruption is triggered by magnetic reconnection which is followed
by a positive feedback between reconnection and outward expansion of
the sheared magnetic flux. However, it is elusive to determine in what
conditions the feedback can be established, and thus to develop prediction methods based on these models is not easy.

On the other hand, the ideal MHD instabilities, mainly the helical
kink instability~\citep[KI,][]{Hood1981, Torok2004, Torok2010,
  Torok2005} and the torus instability~\citep[TI,][]{Bateman1978,
  Kliem2006} are developed based on magnetic flux
rope~\citep[MFR,][]{Kuperus1974, Chen1989, Titov1999, LiuR2020}, which
is a coherent group of twisted magnetic flux winding around a common
axis.
In the MFR-based models, it is assumed that MFR exists in the corona
prior to eruption. Then the pre-existing MFR is slowly driven to an
unstable regime by motions in the lower atmosphere (the photosphere),
and is triggered to erupt through either KI or TI. The twist degree of
the MFR itself and the decay index of the MFR's strapping
field~\citep{Torok2005, Torok2007, Kliem2006} are two critical
parameters which control the KI and TI, respectively, and
instabilities occur when they exceeds some values, i.e., the
instability thresholds. Thus, the controlling parameters of the
instabilities can be used potentially in forecasting solar eruptions,
which is a unique advantage of the ideal MHD models, and to this end,
it is crucial to identify the MFRs in the coronal magnetic field and
to precisely determine the thresholds of the instabilities.

However, there appears to be no single value for the thresholds, and
different studies, from either analytic, numerical or even
laboratory investigations, often give out discrepant results. For
instance, the KI threshold $T_w$, which measures the winding number of
magnetic field lines around the MFR's axis, is found to range from
$\sim 1.25$ turns to $\sim 2.5$ turns in previous analytic and
numerical investigations~\citep{Hood1981, Baty2001, Fan2003,
  Torok2003, Torok2005} because it depends on the
details of the MFR, such as the line-tying effects by the photosphere,
the aspect ratio of the MFR (i.e., ratio of the rope's length to its
cross-section size), and the geometry of the axis. The TI controlling
parameter, i.e., the decay index $n$, refers to the spatial decreasing
speed of the MFR's overlying magnetic field that straps the
MFR. 
A series of theoretical and numerical studies~\citep{Kliem2006,
  Torok2007, Fan2007, Aulanier2010, Demoulin2010, Fan2010,
  Zuccarello2015} show that the TI threshold of decay index has
typical values in the domain of $1.1 \sim 2$, while some laboratory
experiments suggest that it might be much lower as being
$0.65 \sim 1.1$~\citep{Myers2015, Alt2020}, depending mainly on the
ratio of the apex height and footpoint half-separation.

Furthermore, due to the intrinsic complexity of the magnetic field in
the solar corona, the configuration of MFRs can be very different from
case to case, which may not be fully characterized by the KI and TI
theories that are derived based on relatively simplified or idealized
configuration in the previous analytic, numerical and laboratory
investigations. In some cases, if the filaments have significant
rotational motions, they tend to be associated with failed eruptions
despite the decay index satisfying the threshold for torus
instability~\citep{ZhouZJ2019}. Recently, statistical studies based on
magnetic field extrapolation have been employed to study
configurations of coronal MFRs focusing on the two controlling
parameters, in the hope of finding more realistic thresholds that can
be used for prediction and to differentiate the types of eruptive and
confined flares. Using a nonlinear-force-free field (NLFFF)
extrapolation code developed by~\citet{wiegelmann2004} based on the
optimization method, \citet{JingJ2018} surveyed the pre-flare coronal
magnetic field for 38 major flares, and they found that only the TI
plays an important role in distinguishing the different types of
flares, with the threshold of decay index found to be $\sim 0.75$,
which is much lower than the canonical value of
$1.5$~\citep{Kliem2006}. The twist numbers show no systematic
difference between the confined and eruptive events. Then,
\citet{DuanA2019} performed a similar survey of 45 major flares with
the pre-flare force-free magnetic field reconstructed by another NLFFF
extrapolation method, the CESE--MHD--NLFFF code, which is based on the
MHD-relaxation method~\citep{JiangC2013NLFFF}. They used a more strict
definition of MFR which refers to field lines with winding number of
at least one full turn, and found that $\sim 90$\% of the events
possess pre-flare MFRs but with very different three-dimensional
structures. Quite different from the results in~\citet{JingJ2018}, the
newer study found the KI and TI thresholds to be
$|T_w|_{\rm crit} \sim 2$ and $n_{\rm crit} \sim 1.3$, respectively,
which are close to values derived in theoretical studies, and KI and
TI play a nearly equally important role in discriminating the eruptive
and confined flares.

The two previous studies have only analyzed the pre-flare magnetic
fields. In this Letter, we investigate for the first time the
variations of MFRs, focusing on changes of the two controlling
parameters of KI and TI, from immediately pre-flare to post-flare
states. For comparison, we use the same events studied
in~\citet{DuanA2019} and thus the CESE--MHD--NLFFF code is employed to
reconstruct both the pre-flare and post-flare coronal magnetic
fields. It is found that the existences of post-flare MFR are also
very common, even for the eruptive flares, suggesting that the
pre-flare MFR often does not expel out entirely. Furthermore, a key
difference of the variation of the controlling parameters from before
to after flare is shown in two different regimes of events defined by
the KI and TI thresholds derived in our previous study.  These results
provide a strong constraint for the values of the instability
thresholds, and confirm that the events with pre-flare parameters
exceeding the thresholds are very likely triggered by the ideal
instabilities, while for those with the two parameters below the
thresholds, which account for more than half of all the events, other
mechanisms such as the reconnection-based ones should be considered.


\begin{table*}[htbp]
\footnotesize
  \caption{List of events and parameters of their pre- and post-flare MFRs.}
  \begin{tabular}{cccccccccc}
    \hline
    \hline
    No. & Flare peak time & Flare class  & NOAA AR & Position & E/C$^{a}$ & $T_w$ before$^{b}$ & $n$ before & $T_w$ after$^{c}$ & $n$ after \\
    \hline
    1 & SOL2011-02-13T17:38 & M6.6 & 11158 & S20E04 & E &  0.76 & 0.99 & 0.55 & 1.25 \\
    2 & SOL2011-02-15T01:56 & X2.2 & 11158 & S20W10 & E &  1.52 & 0.98 & 0.86 & 0.69 \\
    3 & SOL2011-03-09T23:23 & X1.5 & 11166 & N08W09 & C & -1.75 & 0.50 & -1.45 & 0.30 \\
    4 & SOL2011-07-30T02:09 & M9.3 & 11261 & S20W10 & C & -0.88 & 0.51 & -0.80 & 1.48 \\
    5 & SOL2011-08-03T13:48 & M6.0 & 11261 & N16W30 & E &  2.45 & 1.40 & 1.68 & 1.14 \\
    6 & SOL2011-09-06T01:50 & M5.3 & 11283 & N14W07 & E &  0.92 & 0.52 & 1.30 & 0.40 \\
    7 & SOL2011-09-06T22:20 & X2.1 & 11283 & N14W18 & E &  1.02 & 1.65 & 0.89 & 0.17 \\
    8 & SOL2011-10-02T00:50 & M3.9 & 11305 & N12W26 & C & -0.92 & 0.42 & -0.92 & 0.20 \\
    9 & SOL2012-01-23T03:59 & M8.7 & 11402 & N28W21 & E & -1.63 & 0.73 & -1.40 & 1.17 \\
    10 & SOL2012-03-07T00:24 & X5.4 & 11429 & N17E31 & E & -2.11 & 0.71 & -1.96 & 0.98 \\
    11 & SOL2012-03-09T03:53 & M6.3 & 11429 & N15W03 & E & -1.17 & 0.73 & -1.38 & 0.61 \\
    12 & SOL2012-05-10T04:18 & M5.7 & 11476 & N12E22 & C & -1.11 & 1.21 & -0.77 & 1.11 \\
    13 & SOL2012-07-02T10:52 & M5.6 & 11515 & S17E08 & E & -1.56 & 0.35 & -1.25 & 0.15 \\
    14 & SOL2012-07-05T11:44 & M6.1 & 11515 & S18W32 & C &  1.14 & -0.41 & 1.14 & -0.03 \\
    15 & SOL2012-07-12T16:49 & X1.4 & 11520 & S15W01 & E &  2.20 & 0.42 & 1.83 & 0.28 \\
    16 & SOL2013-04-11T07:16 & M6.5 & 11719 & N09E12 & E & -1.10 & 0.26 & -2.00 & 0.09 \\
    17 & SOL2013-10-24T00:30 & M9.3 & 11877 & S09E10 & E &  2.00 & 0.56 & 1.50 & 0.42 \\
    18 & SOL2013-11-01T19:53 & M6.3 & 11884 & S12E01 & C &  1.50 & 0.42 & 1.60 & 1.40 \\
    19 & SOL2013-11-03T05:22 & M4.9 & 11884 & S12W17 & C &  3.00 & 0.07 & 2.10 & 0.27 \\
    20 & SOL2013-11-05T22:12 & X3.3 & 11890 & S12E44 & E &  1.35 & 2.72 & 1.10 & 1.57 \\
    21 & SOL2013-11-08T04:26 & X1.1 & 11890 & S12E13 & E &  1.26 & 1.87 & 0.80 & 0.49 \\
    22 & SOL2013-12-31T21:58 & M6.4 & 11936 & S15W36 & E & -2.20 & 1.11 & -1.65 & 0.67 \\
    23 & SOL2014-01-07T10:13 & M7.2 & 11944 & S13E13 & C &  1.65 & 0.21 & 0.88 & 0.39 \\
    24$^{d}$ & SOL2014-01-07T18:32 & X1.2 & 11944 & S15W11 & E &  6.50 & 0.20 & 3.86 & 0.31 \\
    25 & SOL2014-02-02T09:31 & M4.4 & 11967 & S10E13 & C & -1.73 & -0.12 & -1.81 & -0.04 \\
    26 & SOL2014-02-04T04:00 & M5.2 & 11967 & S14W06 & C & -1.90 & 1.03 & -1.68 & 0.96 \\
    27 & SOL2014-03-29T17:48 & X1.1 & 12017 & N10W32 & E &  1.53 & 1.72 & 0.60 & 1.37 \\
    28 & SOL2014-04-18T13:03 & M7.3 & 12036 & S20W34 & E &  2.30 & 1.82 & 1.60 & 0.49 \\
    29 & SOL2014-09-10T17:45 & X1.6 & 12158 & N11E05 & E & -0.85 & 0.17 & -1.31 & 0.16 \\
    30$^{d}$ & SOL2014-09-28T02:58 & M5.1 & 12173 & S13W23 & E & -2.76 & 1.96 & -2.00 & 0.57 \\
    31 & SOL2014-10-22T14:28 & X1.6 & 12192 & S14E13 & C & -1.10 & 0.94 & -0.99 & 0.91 \\
    32 & SOL2014-10-24T21:41 & X3.1 & 12192 & S22W21 & C & -1.79 & 0.64 & -1.55 & 0.50 \\
    33 & SOL2014-11-07T17:26 & X1.6 & 12205 & N17E40 & E &  3.55 & 1.21 & 1.60 & 0.51 \\
    34 & SOL2014-12-04T18:25 & M6.1 & 12222 & S20W31 & C &  2.60 & 0.60 & 1.86 & 0.65 \\
    35 & SOL2014-12-17T04:51 & M8.7 & 12242 & S18E08 & E &  0.70 & 0.66 & 0.71 & 0.58 \\
    36 & SOL2014-12-18T21:58 & M6.9 & 12241 & S11E15 & E &  1.09 & 1.49 & 1.00 & 0.27 \\
    37 & SOL2014-12-20T00:28 & X1.8 & 12242 & S19W29 & E &  1.32 & 0.56 & 1.18 & 0.53 \\
    38 & SOL2015-03-11T16:21 & X2.1 & 12297 & S17E22 & E &  2.04 & 1.80 & 1.29 & 0.51 \\
    39 & SOL2015-03-12T14:08 & M4.2 & 12297 & S15E06 & C &  1.10 & 0.72 & 1.30 & 0.99 \\
    40 & SOL2015-06-22T18:23 & M6.5 & 12371 & N13W06 & E & -1.24 & 1.51 & -0.93 & 1.37 \\
    41 & SOL2015-06-25T08:16 & M7.9 & 12371 & N12W40 & E & -2.90 & 0.49 & -2.10 & 0.56 \\
    42 & SOL2015-08-24T07:33 & M5.6 & 12403 & S14E00 & C &  1.04 & 0.33 & 0.73 & 0.36 \\
    43 & SOL2015-09-28T14:58 & M7.6 & 12422 & S20W28 & C & -1.25 & 1.20 & -1.25 & 1.06 \\
    44 & SOL2017-09-04T20:33 & M5.5 & 12673 & S10W11 & E & -1.43 & 1.09 & -1.45 & 0.49 \\
    45 & SOL2017-09-06T12:02 & X9.3 & 12673 & S09W34 & E & -1.80 & 1.72 & -1.10 & 0.22 \\
    \hline
  \end{tabular}
  \tablenotetext{a}{E--eruptive, C--confined.}
  \tablenotetext{b}{before flare--based on the last available magnetogram for at least 10 minutes before the flare $GOES$ start time.}
  \tablenotetext{c}{after flare--based on the first available magnetogram at least 1 hour after the flare peak time.}
  \tablenotetext{d}{Event 24 occurred between NOAA ARs 11944 and
    11943, and event 30 occurred between ARs 12173 and 12172.}
  \label{tab:event_list}
\end{table*}

\section{Data and Method}
\label{sec:method}
We have investigated 29 eruptive flares (above \emph{GOES}-class M5)
and 16 confined ones (above M3.9) observed by the Solar Dynamic
Observatory~\citep[{\SDO},][]{Pesnell2012} from 2011 January to 2017
December, which are listed in Table~\ref{tab:event_list}. All the 45
events occurred within $45^{\circ}$ in longitude of the disk center,
and most of them are active region (AR) flares ~\citep[see more
details of the criterion for selecting events in][]{DuanA2019}.  Based
on the vector magnetograms from the Helioseismic and Magnetic
Imager~\citep[HMI,][]{Hoeksema2014} onboard {\SDO}, the
CESE--MHD--NLFFF code~\citep{JiangC2013NLFFF} was used to reconstruct
a pre-flare 3D coronal magnetic field and a post-flare one for each
event. For the pre-flare reconstruction, we used the last available
data for at least 10 minutes before the flare start time to avoid the
possible artifacts introduced by the strong flare emission, while for
the post-flare field, we used the first available magnetogram at least
1 hour after the flare peak, a time when the post-flare field relaxed
to nearly MHD equilibrium but without significant further stressing by
the photospheric motions. The vector magnetograms are obtained from
the Space-weather HMI Active Region Patch~\citep[SHARP,][]{Bobra2014}
data product.

Following the same procedure introduced in~\citet{DuanA2019}, we first
search for MFRs in the full reconstruction volume based on the
volumetric distribution of the magnetic twist number
$T_w$~\citep{Berger2006}, which can provide a good approximation of
the winding turns of two infinitesimally close field lines about each
other. $T_w$ is defined for closed field lines (whose two footpoints
are anchored in the photosphere) as \begin{equation}\label{Tw}
  T_w=\int_L \frac{(\nabla \times \vec B)\cdot \vec B}{4\pi B^2} dl
\end{equation}
where the integral is taken along the field lines. We calculate $T_w$
on grid points with a resolution of 4 times of the original data, and
define the MFR as a coherent group of magnetic field lines with $T_w$
of the same sign and $|T_w|\ge 1$. Thus an MFR approximately has all
field lines with winding around the axis at least one full turn, which
is a strict definition. Moreover, the MFRs as found with the $T_w$
distribution are further selected by restricting them within the flare
sites, e.g., through comparing with \emph{SDO}/AIA observable
filaments and flare ribbons in the flare site. One may refer
to~\citet{DuanA2019} for more information about the definition and
selection of MFRs. Both~\citet{LiuR2016} and~\citet{DuanA2019}
suggested that the maximum value of $|T_w|$ can be regarded as a
reliable proxy of the MFR axis, and they also pointed out that the
$|T_w|_{\rm max}$ has a very sensitive association with KI occurring
in flares. Thus we employ the $|T_w|_{\rm max}$ as the KI controlling
parameter .

As aforementioned, TI is controlled by the decay index of the
strapping field of the MFR. In many literatures, the decay index is
simply defined as along the vertical or radial direction. However,
nonradial eruptions are frequently observed in filament
eruptions~\citep{McCauley2015}. Thus, in this Letter, the decay index
is calculated along an oblique line matching the direction of the
eruption. In a configuration where the MFR significantly deviates from
the vertical direction, we use a slice cutting through the middle of
the rope axis (often at its apex, marked as P) perpendicularly, and
the slice intersects with the bottom polarity inversion line (PIL) on
the point marked as O. OP is the oblique line directing from O to
P. The strapping magnetic field is extrapolated from the $B_z$
component of the photospheric magnetogram with the potential field
model. Then, it is decomposed into three orthogonal components
$B_{\rm e}$, $B_{\rm p}$ and $B_{\rm t}$, where $B_{\rm e}$ is along
OP, $B_{\rm p}$ is perpendicular to the OP on the slice, and
$B_{\rm t}$ is perpendicular to the slice~\citep[one can refer to
Figure 2 in][]{DuanA2019}. Among them, only the cross product between
the current of the rope and the poloidal flux $B_{\rm p}$ can produce
an effective strapping force directing P to O, since $B_{\rm t}$ is
parallel to the current of the rope, and the cross product of the
current with $B_{\rm e}$ produces a force parallel to $B_{\rm p}$
which can only control the eruption direction, besides, $B_{\rm e}$
often has a small value. Thus, to be more relevant, we define the
decay index $n$ as
\begin{equation}\label{Di}
  n = \frac{d \log(B_{\rm p})}{d \log(r)}
\end{equation}
where $r$ is the distance pointing from O to P.

\begin{figure*}[htbp]
  \centering
  \includegraphics[width=0.9\textwidth]{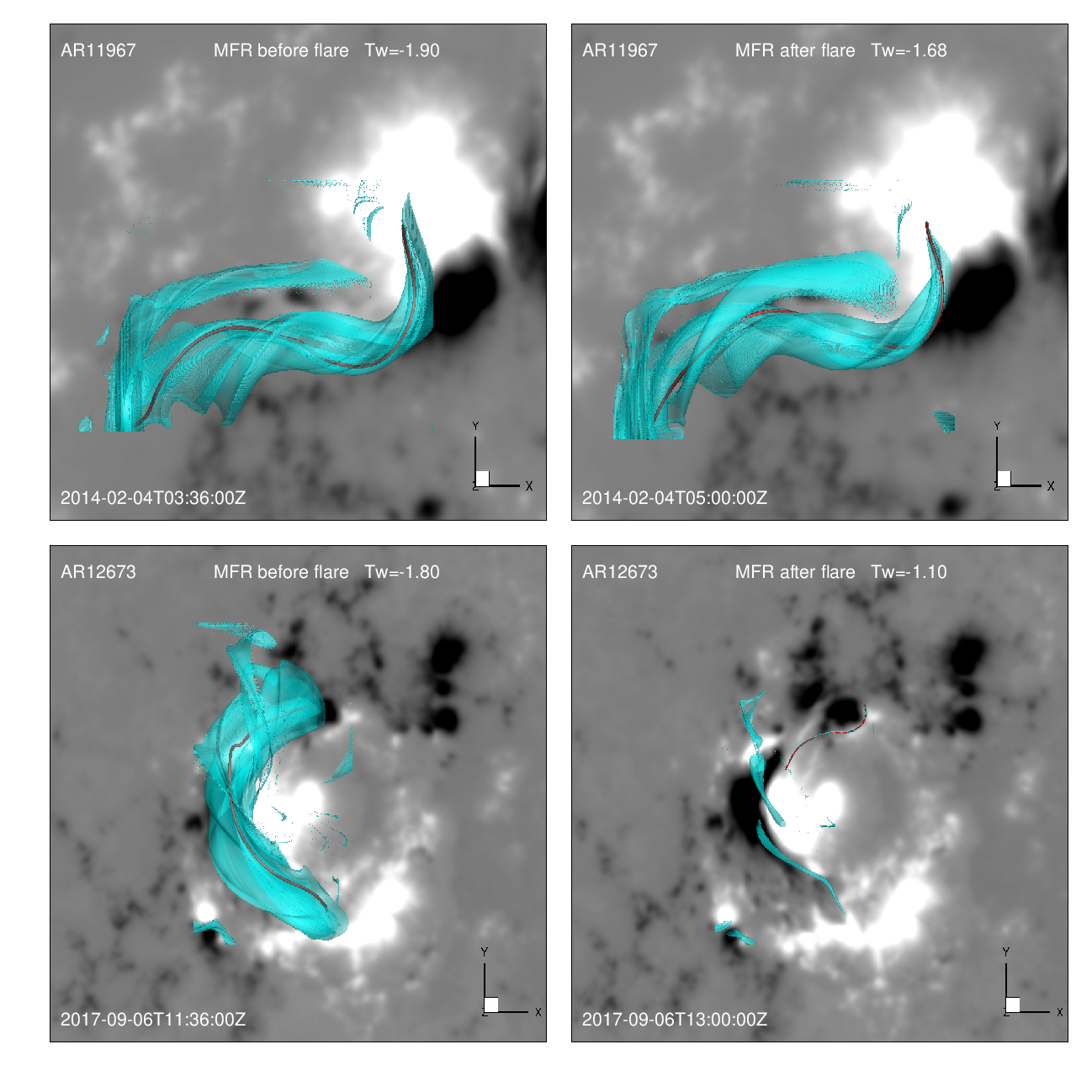}
  \caption{Two examples showing the MFRs before (left) and after
    (right) flares. The top panels show the results for an M5.2
    confined flare (event 26) on 2014 February 4 in AR~11967. The
    bottom panels show results for an X9.3 eruptive flare (event 45)
    on 2017 September 6 in AR 12673. In all the panels, the cyan,
    transparent 3D object is the isosurface of the $|T_w|$ = 1,
    showing the surface of MFRs. Embedded in the isosurface is axis of
    the MFR which is denoted by the red, thick line.  The MFR axis
    possesses the maximum of $|T_w|$ which is denoted on each
    panel. The background images show the magnetic flux distribution
    on the photosphere with the positive flux colored in white and
    negative flux in black.}
  \label{MFR_example}
\end{figure*}

\begin{figure*}[htbp]
  \centering
  \includegraphics[width=\textwidth]{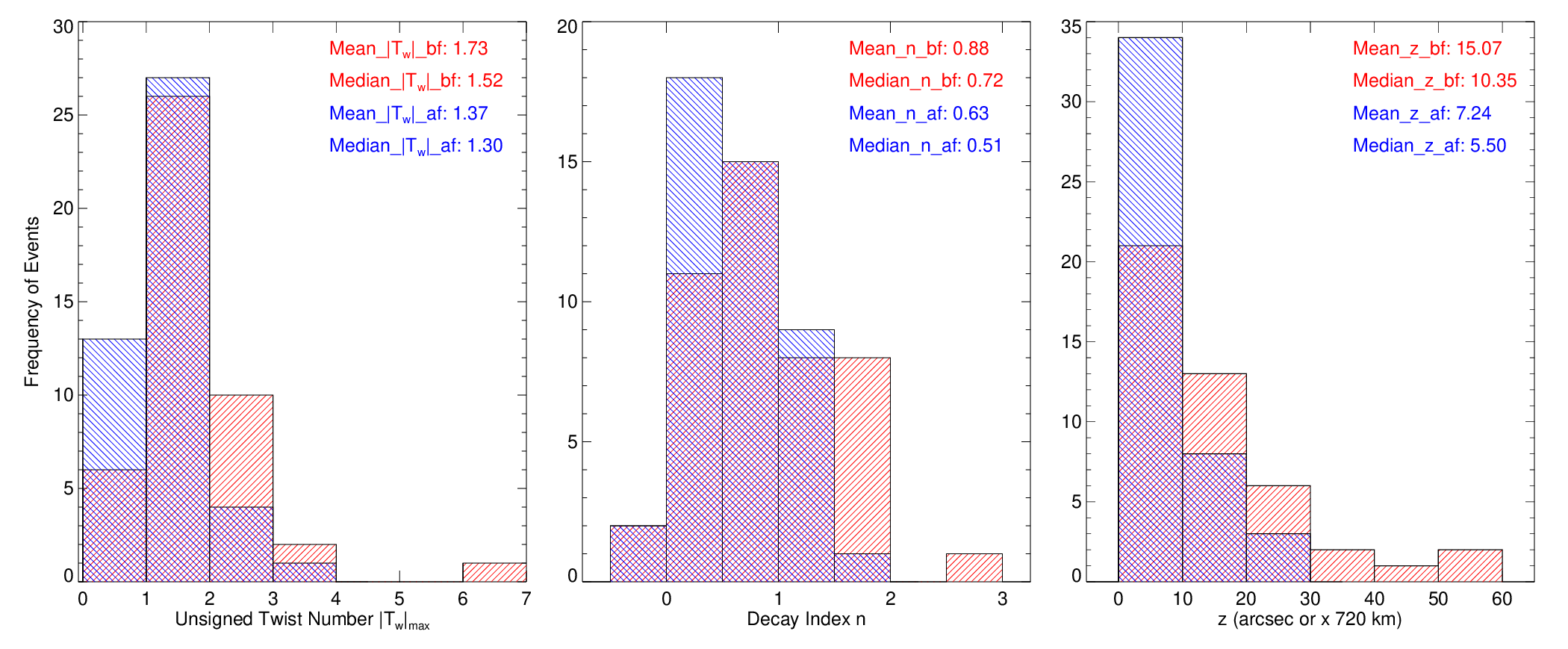}
  \caption{Histogram of the twist number $|T_w|_{\max}$, the decay
    index $n$ and the apex height of magnetic field rope axis
    $z_{\max}$. The red and blue boxes denote before and after flares,
    respectively. The average and median values for all the parameters
    are denoted in each panel.}
  \label{histogram}
\end{figure*}

\begin{figure*}[htbp]
  \centering
  \includegraphics[width=\textwidth]{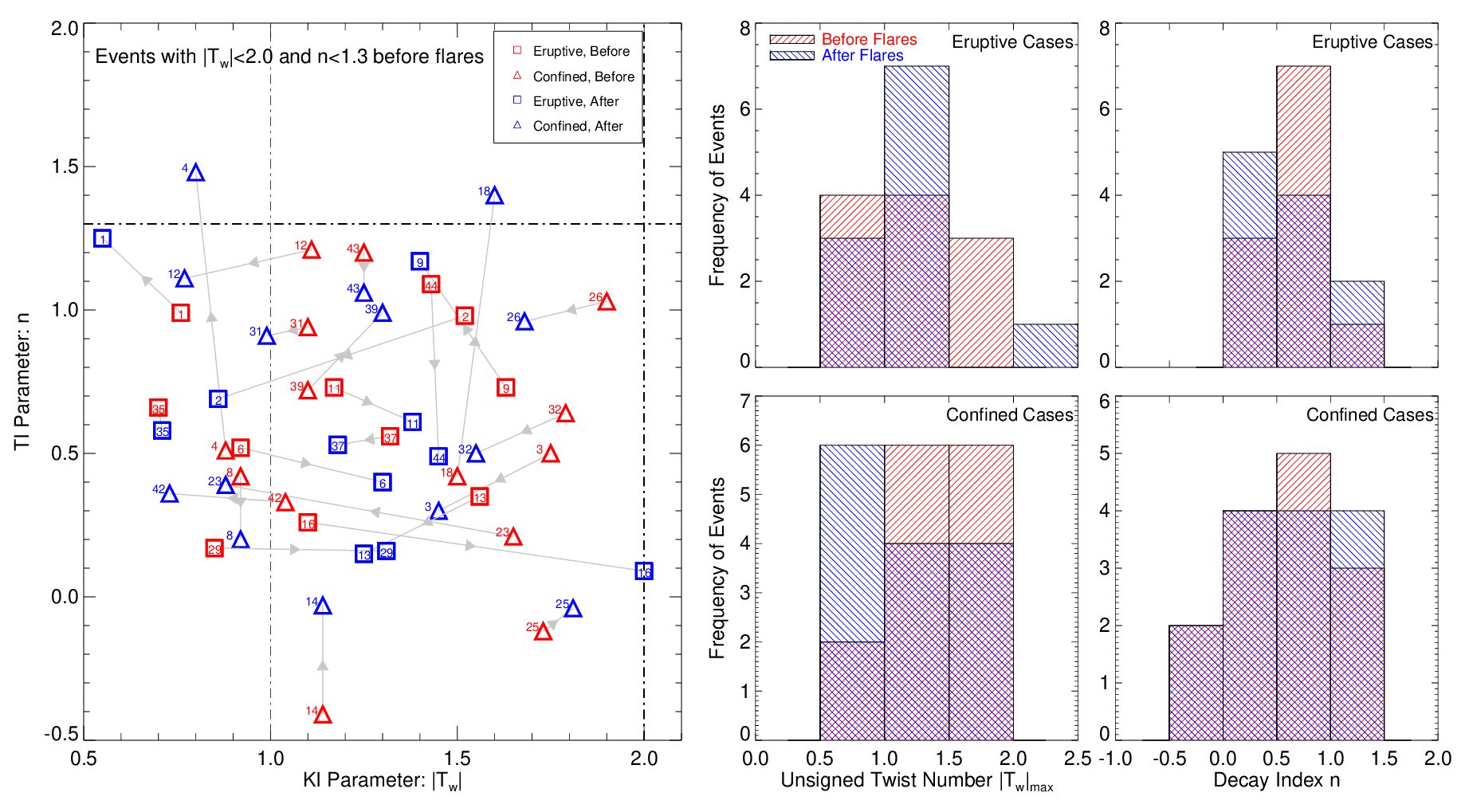}
  \caption{Left: Scatter diagram of decay index $n$ (TI parameter)
    vs. \Twm (KI parameter) for events with $|T_w|<2.0$ and $n<1.3$
    before flares. The boxes (triangles) denote eruptive (confined)
    flares, while red (blue) color denotes before (after) flares, and
    number of the events is marked within or neighboring to the boxes
    and triangles; also see the gray arrows pointing from before to
    after the flares. The thick dashed lines show $|T_w|=2$ and the
    $n=1.3$, i.e., the empirically derived thresholds, while the thin
    dashed line show $|T_w|=1$, below which the MFR does not
    exist. Right: Histogram of the twist number $|T_w|_{\max}$ and the
    decay index $n$ before and after flares for the two types of
    flares.}
  \label{n_and_tw_1}
\end{figure*}

\begin{figure}[htbp]
  \centering
  \includegraphics[width=0.49\textwidth]{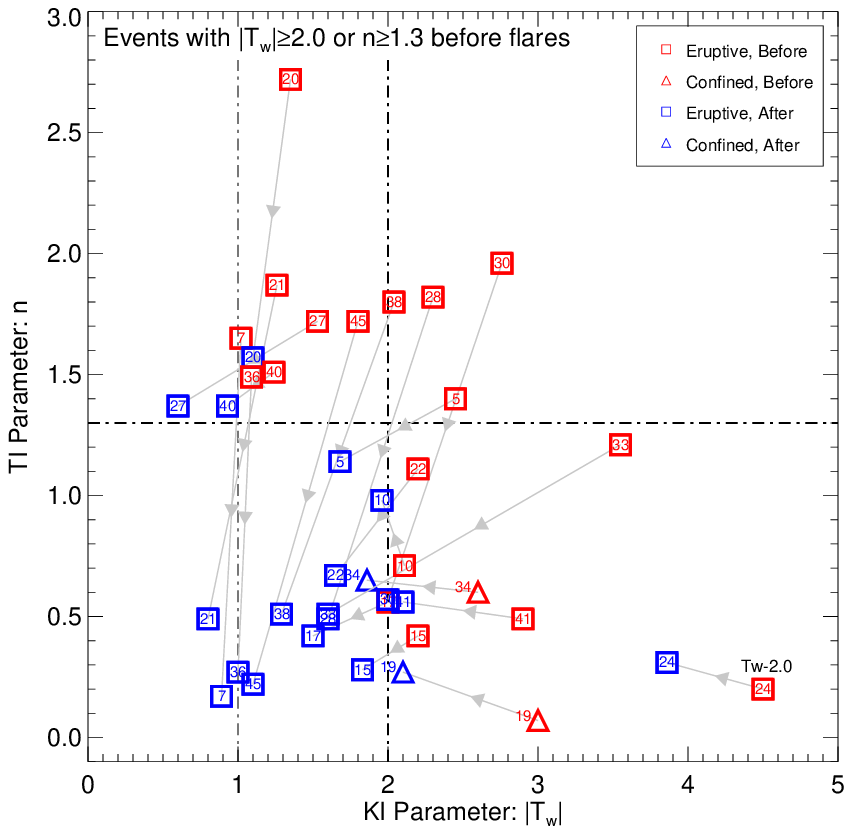}
  \caption{Same as the left panel of \Fig~\ref{n_and_tw_1}, but for
    events with $|T_w| \ge 2$ or $n \ge 1.3$ before flares. Note that
    the pre-flare $|T_w|$ for event 24 is $6.5$, and we show here
    $(|T_w|-2.0)$ for a better display of the plot.}
  \label{n_and_tw_2}
\end{figure}

\section{Results}
\label{sec:res}
Our previous survey \citep{DuanA2019} basing on only the pre-eruption
magnetic fields found that 39 of the 45 events possess pre-flare MFRs
and suggested that the lower limits for TI and KI thresholds are
$n_{\rm crit} = 1.3$ and $|T_w|_{\rm crit} = 2$, respectively, since
90\% of events (11 in 13 events) with $|T_w| \ge 2$ erupted and all
events (11 events) with $n \ge 1.3$ erupted. Thus, it seems that KI
and TI play a nearly equally important role in discriminating the
eruptive and confined flares. Here we focus on how two parameters
change from before to after eruptions.

Before showing the statistical results, we first give two typical
examples of the variation of MFR through two flares, as shown in
Figure~\ref{MFR_example}. The first one is the event 26 in
Table~\ref{tab:event_list}, an M5.2 confined flare on 2014 February 4
in AR 11967. The MFR before the flare has a coherent structure with
peak twist number of $T_w = -1.9$ \citep[see the pre-flare MFR
configurations for all the studied events in][]{DuanA2019}, and after
the flare it shows no evident change with the peak twist number only
slightly reduced to $T_w = -1.68$. The decay index at the apex of the
MFR is $n=1.03$ before the flare and $n=0.96$ after. The small
variation of the MFR through the flare is consistent with the
confining nature of the flare. The second example is shown for event
45, an X9.3 eruptive flare on 2017 September 6 in AR 12673. As shown
in Figure~\ref{MFR_example}, before the flare, there is a thick, long
MFR with peak twist number of $T_w = -1.8$, and it displays a good
coherence and runs roughly along the main PIL of the AR. After an X9.3
flare, the MFR disintegrated, leaving a much thinner MFR with rather
weak peak twist of $T_w = -1.1$, indicating that most (but not all) of
the pre-flare twisted flux of the MFR erupted during the flare in this
case. The remaining MFR after the flare is much lower than the
pre-flare one, and thus the decay index $n$ drops from $1.72$ to
$0.22$. This is a clear example in which both the twist number and the
decay index of the MFR decrease significantly through the eruption.

In Table~\ref{tab:event_list}, we list the parameters for all the
events including the pre-flare peak $T_w$ and $n$ and their post-flare
values. \Fig~\ref{histogram} shows histograms for the distributions of
the two parameters, $|T_w|_{\max}$ and $n$, as well as the apex height
of MFR axis in all the events before and after flares. As can be seen,
on average all the parameters show decrease through flares; the
average $|T_w|_{\max}$ decreases from $1.73$ to $1.37$; the average
$n$ decreases from $0.88$ to $0.63$, and the average apex height of
MFRs is lowered from $15$~arcsec to $7$~arcsec. We note that, as shown
in the $|T_w|_{\max}$ distribution, a majority of the events (32 in
45, or 71\%, which is comparable to the pre-flare percent of 87\%)
after flare still have MFR using our strict definition (i.e., magnetic
flux with $|T_w|>1$), suggesting that MFR existing after flares is
rather common.


Figures~\ref{n_and_tw_1} (left panel) and \ref{n_and_tw_2} show the
scatter diagrams of decay index $n$ versus $|T_w|_{\max}$ in two
regimes for a better inspection. Specifically, in
Figure~\ref{n_and_tw_1}, we show all the events with
$|T_w|_{\max} < 2.0$ and $n < 1.3$ before flares, which are below the
thresholds ($|T_w|_{\rm crit}$ and $n_{\rm crit}$) empirically derived
in our previous survey~\citep{DuanA2019} and thus are not likely
triggered by the ideal instabilities. There are in total 25 events (14
confined and 11 eruptive) in this regime of parameter space. From the
overall change of the pre-flare values (colored in red) to their
post-flare ones (colored in blue) of the two parameters, one can see
the events distribute rather randomly in the region bounded by
$0<n<1.3$ and $0.5<|T_w|<2$ without a systematic decrease in either
the twist number or the decay index. There are only two confined
events, i.e., event 4 and 18, in which the post-flare decay index $n$
increases to above the threshold slightly. In the right panels of
\Fig~\ref{n_and_tw_1}, we show the histograms of the two parameters
before and after flares for the two types of events separately. Again
there is no systematic difference between the eruptive and confined
flares, indicating that the two parameters and its changes are unable
to discriminating the type of flares if both of them are lower than
the thresholds. This is consistent with the conclusion in
\citet{DuanA2019} that the triggering mechanisms of the events fall in
this domain of the parameter space are not related with the ideal
instabilities.

In contrast, Figure~\ref{n_and_tw_2} shows the events with pre-flare
$|T_w|_{\max} \ge 2.0$ or $n \ge 1.3$, namely, the ones with pre-flare
KI or TI parameter (in some cases like events 5, 28, 30 and 38 both KI
and TI parameters) exceeding their thresholds. In this domain of
parameter space, we have in total 20 events in which 18 are
eruptive. Very clear and distinct from the distributions in
Figure~\ref{n_and_tw_1} (left panel), we can see that these events
show systematic decrease in either $|T_w|_{\max}$ or $n$ from through
flares. Furthermore, in most of the events (18 in 20, or 90\% of the
events), both parameters after flares decrease to close to or lower
than their thresholds, i.e., most of the blue marks as seen in the
figure concentrate on the lower left quadrant with $n<1.3$ and
$T_w <2$, except event 20 and 24. These results, again, are highly
consistent with and support our previous statistics~\citep{DuanA2019},
i.e., the thresholds for KI and TI parameters ($|T_w|_{max} \ge 2.0$
and $n \ge 1.3$) are reasonable, and these events are very likely
triggered by the ideal instabilities. It is also worthy noting that
the most of the eruptive events have MFR after flares by our strict
definition (i.e., with $|T_w| > 1$), which indicates that the
pre-flare MFR often does not expel out entirely during eruption.

\section{Summary}
\label{sec:sum}

In this Letter, we systematically studied the coronal magnetic field
changes, focusing on MFRs, before and after solar flares for 45 major
flare events, including 29 eruptive ones and 16 confined ones. Using
the CESE--MHD--NLFFF method with \SDO/HMI vector magnetograms as
input, we reconstructed the coronal magnetic fields immediately prior
to and after the flares for all events. Then we searched the MFR for
each event using a strict definition of MFR based on 3D distribution
of magnetic twist number (i.e., MFR must have a coherent groud of
magnetic flux with twist number of $|T_w| \ge 1$ of constant sign),
and found that while most events possess pre-flare MFRs, the
post-flare MFRs are also very common in a high percent (71\%) of the
events. Furthermore, we calculated the two controlling parameters of
ideal MHD instabilities of MFR, i.e., the maximum twist number
$|T_w|_{\max}$, which controls KI, and the decay index $n$ of the
strapping field which controls TI. For all the events, the average
values for the two parameters show decrease from before to after
flares, with the average $|T_w|_{\max}$ decreases from $1.73$ to
$1.37$ and the average $n$ decreases from $0.88$ to $0.63$, indicating
of magnetic twist releasing and MFR height decreasing through flares.

A key difference of the variation of the two parameters from before to
after flare is shown in the two different regimes defined by the KI
and TI thresholds derived in our previous study~\citep{DuanA2019},
which are $|T_w|_{\rm crit}=2$ and $n_{\rm crit}=1.3$,
respectively. For the events with both parameters before flares below
their thresholds, most of them after flares are also lower than their
thresholds, without systematic change of the parameters found from
before to after the flares, and the two parameters and their changes
are unable to discriminating the eruptive or confined type of
flares. While for the events with any of the two parameters exceeding
their threshold before flare, there is systematic decrease in either
$|T_w|_{\max}$ or $n$ after flare, and in most of these events, both
parameters decrease to close to or lower than their thresholds. Thus
the pre-flare to post-flare changes of the two parameters confirm our
empirically derived thresholds for KI and TI (i.e.,
$|T_w|_{\rm crit}=2$ and $n_{\rm crit}=1.3$), above which these events
are very likely triggered by the ideal instabilities and can be
eruptive. While for those with the two parameters below the
thresholds, other eruption mechanisms such as the reconnection-based
ones should be considered to understand the triggering of the flares
and the conditions determining the eruptive or confined types. These
results give a strong constraint for the values of the instability
thresholds and also stress the necessity of exploring other eruption
mechanisms in addition to the ideal MHD instabilities.



\acknowledgments

This work is jointly supported by the B-type Strategic Priority
Program XDB41000000 funded by the Chinese Academy of Sciences,
National Natural Science Foundation of China (U2031108) and the
startup funding (74110-18841214) from Sun Yat-sen University.
C.W.J. acknowledges support by National Natural Science Foundation of
China (41822404, 41731067, 41574170, 41531073). Z.J.Z. is supported by
NSFC grant 42004142.  Data from observations are courtesy of NASA
{SDO}/AIA and the HMI science teams. Special thanks to the anonymous
referee for invaluable comments and suggestions that improved the
paper.


\end{CJK*}
\end{document}